\documentclass{article}
\usepackage[english]{enigtex}

\usepackage{pifont} 

\usepackage{nth}

\usepackage{fixltx2e} 

\usepackage{tikz}
\usetikzlibrary{decorations.pathmorphing}
\usetikzlibrary{decorations.markings}
\usetikzlibrary{shapes.multipart}
\usetikzlibrary{intersections}
\usetikzlibrary{calc}
\usetikzlibrary{positioning}

\begin{document}

\newcommand{\HRule}{\rule{\linewidth}{0.5mm}}
\begin{titlepage}
	\linespread{3}%
	\selectfont
	\begin{center}
		\vspace*{\fill}
		\HRule \\[0.5cm]
		{\huge \bfseries General Covariance and Background Independence in Quantum Gravity}\\[0.3cm]
		\HRule
		\vfill
	\end{center}
\end{titlepage}
\linespread{1}%
\selectfont

\begin{abstract}
	The history of general relativity suggests that in absence of experimental data, constructing a theory on philosophical first principles can lead to a very useful theory as well as to ground-breaking insights about physical reality.
	The two related concepts of general covariance and background independence are some of these principles and play a central role in general relativity. A definition for them will be proposed and discussed.
	
	Constructing a theory of quantum gravity -- the sought-after unification of general relativity and quantum mechanics -- could profit from the same approach.
	Two popular research areas of quantum gravity, string theory and loop quantum gravity together with spin foams, are being examined whether they comply with these principles.
\end{abstract}

\setcounter{tocdepth}{2}
\tableofcontents
\newpage

\section{Introduction}

The theory of general relativity is a remarkable achievement in several ways. Not only does it satisfy the experimental constraints with high accuracy, but it is constructed upon deep philosophical insights, as we will see later. 

In quantum mechanics, history took a different path. Unexpected outcomes of experiments forced the view of physical reality into a new shape. Heuristic theories were fit to the observations, leaving huge philosophical and interpretational questions open.
Even today, there is controversy about the interpretation of collapse in quantum mechanics.
Other foundational questions such as ``What is a state?'' and ``What is an observable?'' have been axiomatised meanwhile in different ways, still without universal agreement.

Unifying general relativity and quantum mechanics is a notorious open problem.
It has been recognised\footnote{See \cite[section 1.1.2]{rovelli:qg}.} that in the absence of experimental hints, deduction from philosophical first principles such as background independence and closely related concepts might help in finding a solution of this problem.
The aim of this essay is a thorough discussion of this idea, and finding out in what extent it is present in modern research.\\

In section 2, different notions of background independence will be discussed. Definitions of central concepts such as ``dynamical fields'' and ``background fields'' are emphasized and their relevance explained. It will turn out that background independence in classical field theories is not a complicated issue: It is mostly differing definitions that can lead to confusions.

We will then turn to a discussion of philosophy of general relativity in section 3, in particular the implementation of general covariance and background independence. The question which of the cornerstones of general relativity are particularly important for quantum gravity, will be raised and discussed.

Afterwards, two popular approaches to Quantum Gravity, string theory (section 4) and loop quantum gravity together with spin foam models (section 5), will be examined for their compliance with the notions of background independence that have been developed previously.
A recurring theme will be that background independence is hard to define for quantum theories that introduce radically new concepts of space and time. We will have to augment our definitions from case to case. The question whether a specific approach to quantum gravity is background independent thus becomes much more challenging.

Finally, in section 6 we will compare our gained knowledge from the sections about strings and loops and comment on the importance of this whole discussion. 

\section{Background independence}

The issue of background independence is sometimes only touched briefly in the literature. This is one of the reasons why (to the knowledge of the author) there are no prevalent definitions which are generally agreed upon.
Quite a lot of terms such as ``dynamical'' and ``background'' remain dubious. This is one common source of confusion and sometimes leads to someone claiming that a particular theory is background dependent whereas the proponents do not see a problem because they use a different definition of background dependence.

One of the main goals of this essay is proposing a set of consistent definitions and a discussion of the different meanings one could have in mind when claiming that some theory is background dependent or not.

This notion is indeed relevant, as we will see in section \ref{philosophygr}, using the example of general relativity.
One of its leading ideas was to give up completely the idea of absolute spacetime.
First of all, Newtons absolute space and time, but also specially relativistic spacetime has to be given up.
We will later see that in the light of general relativity, an absolute spacetime of any kind would be interpreted as a ``background''. So one of the leading ideas of general relativity was background independence. This will be made clear in section \ref{gravitationalfield}.

\subsection{Fields}

The idea of a physical field dates back to Faraday, who introduced the picture of flux lines, and was formalised by Maxwell with his famous Maxwell equations of electromagnetism.
Sometimes, when doing particle mechanics, we assume a field to be given and study the particle motion in it. In other situations, we will consider the dynamics of the field itself, like the study of electromagnetic waves. In the latter case, the field is a ``dynamical'' field, while in the former case, the field will be called a ``background'' field. The concrete definition will be given in section \ref{dynbgfields}.

Modern particle physics and gravitation theory tell us that all known fundamental physics is based on (classical or quantum) field theory. Naturally, the question arises whether all observable fields are dynamical. Theories where this is the case will be called ``background independent''.

\subsubsection{Classical fields}

Let us first set up a vocabulary of the objects we will be working with.

A rather restrictive definition of a field in classical physics would be:
\begin{defn}
	A classical (scalar) \textbf{field} is a function $f$ from spacetime $M$ to a vector space $V$.
\end{defn}
The vector space $V$ could be the reals, the complex numbers or even a Lie algebra, as in Yang-Mills-theories.

Normally, we will want to be less restrictive extend the definition to sections of vector bundles.\footnote{The previous definition is then recovered as a special case: A function from $M$ to $V$ is a section of the trivial bundle $M \times V$.}
This way, we include all tensors such as vector fields and differential forms in the definition.
(One other way of generalisation would be to take into account arbitrary manifolds for $V$ instead of only vector spaces, as in nonlinear sigma models.)\\

It is worth emphasizing that these definitions hold only in classical physics.
A ``quantum field'' cannot be defined as easily. We have to be careful when trying to generalise concepts from classical background independence to quantum physics, where we might not have precise mathematical definition at hand.
We might even have to alter our understanding of backgrounds.

\paragraph{Examples}
\begin{itemize}
	\item A potential for a classical point particle is a real-valued function \[V: M \to \R\]
	\item The classical electromagnetic potential, a $U(1)$-connection can be written as a differential 1-form in a trivialisation: \[A \in \Omega^1(M)\]
	\item The metric tensor from general relativity is a symmetric, non-degenerate tensor of rank 2: \[g \in \operatorname{Sym}(T^*M \otimes T^*M)\]
\end{itemize}

\subsubsection{Actions and Lagrangian densities}

Defining what fields are explained the kinematics of field theories: The states of a classical field theory is just a field.

The dynamics of field theories are usually governed by differential equations, the field equations. The solutions of the field equations are the allowed configurations of the field.
All solutions that can't be distinguished in any measurement are considered to form one physical state that occurs in the real world.
A state thus is an equivalence class of configurations.
If a state contains more than one configuration, there is gauge redundancy. A transformation that maps any configuration onto a configuration of the same state is called a gauge transformation.

There are different ways to arrive at field equations. One can postulate one, based on an educated guess (like the Einstein equations) or experimental data (electromagnetism before Maxwell). One could also set up a Hamiltonian density and then take Hamilton's equations.

Probably the most common way nowadays is to define a Lagrangian density and consider the Euler-Lagrange equations which follow from varying the action.
This has the advantage that gauge symmetries of the Lagrangian density become symmetries of the equations.
In this discussion, we will restrict ourselves to the discussion of the Lagrangian approach since all the examples later on make use of it.\\

The action $S$ is defined as the integral of the Lagrangian density $\La$ over all of spacetime:
\[S[f_1, f_2,\ldots] = \int_M \La(f_1, f_2,\ldots, g_1, g_2,\ldots)\]
It is a functional of the fields $f_1, f_2,\ldots$ and of its derivatives. We may use other fields $g_1, g_2,\ldots$ to define the action.

In order to make this integral well-defined, we must insist that \\$\La(f_1, f_2,\ldots, g_1, g_2,\ldots)$ is a differential top-form, since only differential forms of degree of the dimension of spacetime can be integrated.\footnote{On spacetimes with an orientation and a metric (such as ordinary solutions of the Einstein equation) carry a canonical volume element $\sqrt{\pm \lvert\det g\rvert}\diff x$ and every function on spacetime can be integrated by first multiplying with the volume element and then integrating as usual.}

\subsubsection{Dynamical fields and background fields}

\label{dynbgfields}

We are now ready to state the definition of dynamical and background fields in classical field theory.
\begin{defn}
	A field $g$ is called \textbf{non-dynamical} or \textbf{background field} if the action depends on it, but is not a functional of it.\\
	Other ways of stating this are:
	\begin{itemize}
		\item Changing $g$ does not mean changing the field configuration, but rather changing the theory or at least a (high-dimensional) parameter of the theory.\footnote{Compare particle mechanics with a harmonic oscillator potential to particle mechanics in a Coulomb potential. These are not different states of the same theory, but rather different theories.}
		\item The action depends on $g$, but $g$ is not being varied with respect to it. This implies that there are no field equations\footnote{We may of course only consider background fields $g$ which satisfy certain field equations, but that is not interpreted to be part of the theory we are considering.} for $g$. (It therefore doesn't have dynamics, hence the name ``non-dynamical''.)
	\end{itemize}
	Otherwise, if the action is a functional of $g$, it is called a \textbf{dynamical field}.\\
	
	An action is called \textbf{background independent} if it can be defined without background fields, that is, all fields are dynamical.\\
	
	A field theory defined by an action is called background independent if its action is background independent.
\end{defn}
An important feature of a background, as emphasized in \cite{rozali}, is that in principle, it can be changed in a physical process, but our description does not allow for such a change.
We then assume that there is a more \emph{fundamental} theory where the former background field also has field equations and becomes dynamical.\\

Let us call a theory ``fundamental'' when it is true to the best of our knowledge. A fundamental theory has not been falsified by any evidence and consistently describes all of the phenomena it attempts to describe.

\paragraph{Examples}

A generic case is a field $f$ on a fixed Lorentzian manifold with a Lagrangian density function $L$. With $g$ being the metric, it has the following Lagrangian density: \[\La(f,g) = L(f) \; \sqrt{- \det g} \; \dif^n x\]
The action then is: \[S[f] = \int_M \La(f,g)\]

This action clearly is background dependent: It is varied with respect to $f$, but not with respect to $g$. The metric is a background field. Different values of $g$ correspond to different situations we could study the dynamics of $f$ in, not to different states of the same theory.\\


The prototypical example of a background independent field theory is general relativity.
The Einstein-Hilbert action 
is the integral of $\La(g) = R \sqrt{- \det g} \dif^n x$ with the curvature scalar $R$. It will be discussed in more detail in sections \ref{philosophygr} and \ref{diffeocovarianceqg}.\\
Again, the Lagrangian density depends on $g$, but this time it is background independent since $g$ is the dynamical field of the theory and we are interested in field equations for $g$, which will later turn out to be the Einstein equations.\\

We note that the difference between background dependent and independent may come from our interpretation. A background independent theory is one where we are studying the dynamics of all fields simultaneously. We still have not decided whether we are convinced that all fields we ever encounter should in fact be dynamic.

\subsection{Perturbative definition $\implies$ background dependence}

\label{perturbative}

In the quantum field theory community and especially amongst string theorists, ``background independent definition'' is often used synonymously with ``nonperturbative definition''. Let us argue that the latter is a special case of the former, which means that any perturbative definition of a field theory is a background dependent definition.\\

The action of a field perturbation $f$ around a fixed field $\phi_0$ is: \[S_{\phi_0}[f] := S[\phi_0 + f]\]
Obviously, $\phi_0$ is a background field which enters the definition of the action but is not dynamical since it is fixed.

If we want a background independent theory, we cannot allow perturbation theory in the fundamental definition which is believed to be unrestrictedly true.
The converse statement is not necessarily true: Some theory might not be defined with the help of perturbation around a background, but there might be another background it depends on, for example the metric. This is a common source of misunderstandings.
The terms ``background independent'' and ``nonperturbative'' therefore should not be confused without clarification.\\

In classical field theory, this is quite tautological. We even assumed the existence of a nonperturbative action $S[\phi_0 + f]$. It is generally not necessary to define a classical theory with the help of perturbation theory.
Defining a theory perturbatively does not necessarily render it background dependent.
There might be several ways of defining it.
If there is a fundamental, background independent definition and we show that we can derive the other definitions from it, the full theory is background independent.
However, in quantum field theory, the situation might become more subtle:

We might want to do a path integral over small perturbations $f$ around a vacuum $\phi_0$. We therefore use the background dependent action $S_{\phi_0}$. In order to handle the integral, we might have to calculate power by power in a coupling constant, relying on the action being analytical around $\phi_0$ and the higher powers vanishing fast enough.

In the Hamiltonian picture, we might define the (quantum) state space as excitations around a vacuum field configuration $\left|\phi_0\right>$.

Either way we introduced a background field to define a quantum field theory. The resulting quantum field theory thus is background dependent.
There is no such easy reformulation like in the classical case: The background field $\phi_0$ remains classical, whereas the perturbation $f$ becomes a quantum field.

In a background independent field theory however, we can apply perturbation theory only \emph{after} having defined the theory non-perturbatively.\\

It has to be remarked that there can be another reason for a perturbative definition not to be fundamental: This is the case if the power series does not converge. 

A typical example is interacting quantum field theory in four dimensions.
The Fock space is defined by plane waves of the noninteracting theory, but the perturbation series is defined in powers of the interaction terms.
This perturbation series usually does not converge\footnote{This is true even after renormalisation.} at all. We can only hope that for small coupling constants, the first few terms give a good approximation and the higher terms must not be included for some reason yet to understand.
Our definition is not fundamental in that case, it is at most an ``effective'' description.
We are in such a situation in quantum electrodynamics, where experiments agree to enormous precision with the first terms of the perturbation series, but the series is known not to converge.


\subsection{Diffeomorphism covariance $\implies$ background independence}

Our next aim is to clarify one further confusion: The synonymous usage of the terms ``diffeomorphism covariant'' and ``background independent''. In our notation, the former will be a special case of the latter.\\

\subsubsection{Active and passive diffeomorphisms}

A diffeomorphism\footnote{A diffeomorphism $\phi$ is a smooth map of manifolds with smooth inverse. An intuition for it will be supplied in an example. Diffeomorphisms can act on tensor fields on a manifold as well by a map called the pull-back $\phi^*$.} and  of spacetime can be interpreted in (at least) two ways, as a change of coordinates and as a gauge transformation acting on the field configuration.\\

The first interpretation builds on the insight that it is possible to use different coordinates to describe the same physical situation.
For example, we are allowed to solve the classical, prerelativistic equations of motions of a point particle in an inertial frame of reference or in a rotating frame.\footnote{Of course, that description is correct only with fictitious forces.} We could even adopt spherical coordinates and derive equations of motion for the radius and the angular coordinates.

All these are equivalent descriptions of the same physical situation. A clever physicist will choose a description that suits the situation well. It is possible to change from one description to the other because the coordinates are related by a diffeomorphism.\footnote{In some cases, the diffeomorphism is only local. For example around the origin, polar coordinates are not related to cartesian ones by a diffeomorphism.}

Adopting the nomenclature used in \cite{rovelli:qg}, we will call diffeomorphisms interpreted in this way ``passive diffeomorphisms''.\\

If we interpret diffeomorphisms as gauge transformations, the situation is different.
Such a gauge transformation in field theory maps a field configuration into a physically indistinguishable field configuration via a pull-back. So we may change the field configuration (as long as all the observables stay the same), but we also leave the description (e.g. the coordinate system) the same.

We will call diffeomorphisms interpreted in this way ``active diffeomorphisms''.\\

Let us consider a simple example to understand this abstract difference.
Consider a stage with objects on it.

A passive diffeomorphism just amounts to a change of view: We may watch from a seat in the audience and some object may appear on the left. We may watch from the backstage area and the same object appears on the right.

An active diffeomorphism lets us stay at the same observation point but turns the objects on the stage around according to some predefined transformation behaviour.\\

One may think at first that there is no observable difference in these two sorts of diffeomorphisms. But there is a difference in if we recognise that there is a background present, the stage.

Imagine equidistant markings on the stage. 
While a passive diffeomorphism will not allow a perceivable change in quantities independent of the observation point, an active diffeomorphism will. In that case, the theatre play is not invariant under the action of an active diffeomorphism. The reason for this is the stage, a background.\footnote{If there were no stage at all, 
we could not tell the difference between an active and a passive diffeomorphism.
This is very hard to visualise: From our everyday experience, we always imagine space containing the objects. But it will turn out that exactly this space also is a background.} 
The theatre play is not invariant because the active diffeomorphism acts on the objects (the dynamical part), but not on the background.

From our understanding of the previous arguments and the example, we can now extract the definition of active and passive diffeomorphisms in classical field theory:
\begin{description}
	\item[Passive diffeomorphisms] act on \textbf{all} fields (dynamical and non-dynamical)
	\item[Active diffeomorphisms] act \textbf{only} on \textbf{dynamical} fields
\end{description}

\subsubsection{Diffeomorphism covariance, or general covariance}

\label{diffeocovariance}

Invariance under passive diffeomorphisms is nearly a tautology in classical field theory. If all fields are smooth and the theory is formulated with the help of a Lagrangian density, we can just transform some solution of the field equations and we automatically know that the transformed solution will solve the transformed field equations.

Invariance under active diffeomorphisms is a severe constraint. This is the actual, physically interesting constraint we are interested in, therefore we will call it \textbf{diffeomorphism invariance}, or \textbf{general covariance}.
It implies that any solution of the field equations can be transformed and still satisfies the same, untransformed field equations.
This implies that any coordinate system is appropriate for solving the field equations since they look the same everywhere.

A prominent example of this is general relativity: If the metric $g$ is a solution to the Einstein equations and $\phi$ if a diffeomorphism, then $\phi^* g$ (the pull-back) also is a solution to the Einstein equations.

The background fields normally appear in the field equations and therefore have to stay untransformed under an active diffeomorphism.
If the background fields are in any way coupled to the dynamical fields, different values of the background fields at two distinct points in spacetime will most likely make the field equations look different at these points. In any sensible field theory with background fields, a solution to the field equations will not be mapped onto a solution again by a diffeomorphism that interchanges these two points.

We must infer that any background field in a diffeomorphism covariant theory has to be a scalar constant.

There are two possibilities now:
\begin{itemize}
	\item We could choose to interpret the background field, if it is a scalar, as an actual constant of nature, and not a field at all. All occurences of the field then have to be replaced by this constant of nature.
	
	If the field under question is not a scalar, we must restrict ourselves to a specific set of coordinate systems and interpret the components of the field as scalar fields, which then have to be constant. Diffeomorphism invariance then is not a gauge symmetry of the theory.
	This is the approach that was chosen for the metric field before general relativity. 
	The specific set of coordinate systems then are inertial frames.

	\item We might recognise that the background field in question should be interpreted as a dynamical field.
	
	General relativity will turn out to be an excellent example of this case.
	The metric has to change its role from a background field to a dynamical field here.
\end{itemize}
In either possibility, if it is possible to retain diffeomorphism invariance, the theory in its new interpretation is background independent.
We see again that the difference between background dependence and independence can sometimes be purely interpretational: It matters if we interpret something as a field or as a constant, and also whether we prescribe dynamics to it or not.

\section{Classical and quantum gravity}

We have some vocabulary of background independence in classical field theory. However, a caveat has to be pointed out here: When discussing quantum gravity, much of the vocabulary will loose its meaning, because quantum gravity is by nature not classical.

Already quantum field theory requires us to rethink our understanding of background independence, as done in section \ref{perturbative}.
We can already say that the fundamental definition of quantum gravity will be a nonperturbative one because situations with extreme changes in geometry such as black holes or the big bang will have to be describable.

If we approach such an unestablished field as quantum gravity, we have to expect some of our definitions not to be applicable anymore.
In some approaches such as loop quantum gravity, we will end up with a discrete structure instead of a smooth manifold for spacetime. According to the Anti de Sitter/Conformal Field Theory (AdS/CFT) - conjecture, spacetime is just a holographic effect and the fundamental description is formulated on the lower-dimensional boundary of spacetime.
Under such circumstances, we will have to stay alert for possible redefinitions of background independence. Ambiguities may occur and make the discussion of background independence nontrivial.

\subsection{The research area quantum gravity}

Sometimes, quantum gravity is being treated as an established theory rather than an area of active research with many completely different approaches. This is of course a misjudgement.
The best current theory to describe gravity is general relativity, a classical, background independent field theory.
Until now, no fully satisfying theory has been found that has general relativity as some sort of classical limit and is compatible with quantum mechanics.
Most candidate theories lack mathematical rigor, a fundamental definition beyond the regime of small perturbations or a proof of the correct classical limit.

Such a theory has the challenge of following the gigantic footsteps set by general relativity. Einstein's remarkable theory 
changed our understanding of spacetime profoundly. A satisfying theory of quantum gravity is expected to change our understanding once again, with the same importance.

More specifically, to become widely accepted it should explain how the geometry of spacetime is quantised and also - sometimes underestimated - how quantum particles source the ``quantum gravitational field''.\footnote{By this, we mean the yet unknown quantum concept that will replace the classical dynamical spacetime.}

\subsection{Philosophy of general relativity}

\label{philosophygr}

In his struggle of defining general relativity, Einstein didn't have concrete experimental hints.\footnote{It is sometimes claimed that the already observed anomalous perihelion shift of mercury was an important hint, which is not accurate. The perihelion shift didn't specific lead to any feature of the theory, it just pointed out that the Newtonian theory of gravity was not applicable here anymore, which Einstein already knew from theoretical deduction. Rather, the perihelion shift provided a valuable test of general relativity afterwards.}
Rather, he deduced the theory from the known nonrelativistic limit (Newtonian gravitation) and the profound philosophical insight of general covariance, which we paraphrased as ``diffeomorphism covariance''.

To understand what exactly the insight was, let us take a small detour through Newtonian mechanics and special relativity.\footnote{This discussion is covered in great detail in \cite[section 2.2]{rovelli:qg}. Only the central points are being sketched here.}\\

In Newtonian mechanics, the equations of motion for the position $\vec x$ of a point particle of mass $m$ are: \[m\ddot{\vec x} = \vec F\]
These equations of motion satisfy the Galilean principle in that they do not change under a Galilean transformation.\footnote{This is a transformation where the new coordinates $\vec y$ are defined as $\vec y = R \vec x + \vec v \cdot t + \vec a$, where $R$ is a constant rotation matrix, $\vec v$ is a constant velocity and $\vec a$ a constant translation vector. It is easy to see that $\ddot{\vec y} = R\ddot{\vec x}$ and therefore the equations of motion $m\ddot{\vec y} = R\vec F$ hold for $y$. Thus, Newtonian mechanics is Galilei-covariant, as demanded by the Galilean principle. (At this point, ``covariant'' is more suitable than ``invariant'' since the right hand side of the equation indeed transforms nontrivially, but accordingly.)}
One could say that Newtonian space is not necessarily completely absolute since in the coordinate system that $x$ is situated in, neither the absolute origin nor absolute rest can be determined.
However, Newton still assumes (with unease) that \emph{acceleration} is an absolute concept. There is a preferred reference frame which can be determined up to a Galilean transformation.

In special relativity, Einstein reasserts the Galilean principle and shows that it is possible to unify it with electromagnetism if one gives up the notion of absolute time.
However, he is not quite content with it, as it seems to weak to him.
He postulates that the fundamental equations of motion must be covariant under transformations to arbitrary reference frames, that is, under diffeomorphisms.
This is justified by his inobvious resolution of the famous ``hole argument''.

\subsubsection{The hole argument}

\label{holeargument}

Einstein's hole argument is the attempt to negate the question whether the equations of the gravitational field should be covariant.
Imagine the following situation: In spacetime, let there be two spacetime points $x$ and $y$.
Let us assume without further specification that the gravitational field at $x$ is very strong whereas there is zero gravity at $y$. If we were to demand diffeomorphism covariance from our field equations, then we could apply a diffeomorphism that interchanges $x$ and $y$.
Suddenly we find the strong field at $y$ and a zero field at $x$.
It seems at a first glance, that such a theory is completely unpredictive as it does not tell us the field value at a given spacetime point.\\

However, there is a subtle flaw in the argument. It appears already in the very first assumption: ``In spacetime, let there be two spacetime points $x$ and $y$.''
Einstein's insight now is: \emph{The idea of spacetime points on their own does not have any physical meaning.}

Similarly as there is no absolute origin of the coordinate system of Newtonian physics, the radical generalisation is now to demand that there are no spacetime points on their own at all.
We may define spacetime to be a manifold and refer to points of it, but there is no physical significance in them.

One other way to recognise it is the following: We have established in section \ref{diffeocovariance} that not passive diffeomorphisms (which are just a change of coordinate systems of the same region of the manifold), but active diffeomorphisms, which map parts of the manifold on others, are the gauge transformations of diffeomorphism covariance. Consequently, points are transported by these transformations as well.\\

The main lesson we learn from this is a philosophical one: It is consistent to demand general covariance from a physical theory, but we have to accept the huge philosophical conclusion that there is no meaning of spacetime points in themselves.

How can we meaningfully talk about field values at certain points then?
Often, we implicitly choose a certain coordinate system\footnote{One could say we fix a certain gauge.} when referring to a measurement apparatus.
If we say that the apparatus measured a specific field value at a given point in spacetime, we actually define a coordinate system that has the laboratory (or some other chosen object such as the center of our galaxy) as the origin and an arbitrarily chosen moment as zero time.
The point here is that we choose a coordinate system by relating it to \emph{coincidences in spacetime}.

The origin of the coordinate system is not being specified by choosing a point on a manifold but by demanding that, say, the finger of the experimenter touches the internal clock of the measurement apparatus.
We choose the axes of the coordinate system in a convenient way, maybe such that the spatial coordinates of the apparatus stay constant, to simplify calculations. (We could indeed choose any system we like, as we know.)
Also during the actual measurement, the same logic applies:
We do not measure the value at a given point on the spacetime manifold, but we measure through a spacetime coincidence.
It is the coincidence of, say, several electrons following some trajectory defined by the interaction with the electromagnetic field that we interpret as a measured field value.
In this sense, under close enough inspection, measurements are compatible with general relativity.\footnote{Note that we arrive at this conclusion by interpreting the measurement apparatus as a part of the physical system.}

\subsubsection{The gravitational field replaces space and time}

\label{gravitationalfield}
If spacetime is not objective on its own, with respect to what does an accelerated observer accelerate? Clearly, there must be something real about it as acceleration has observable effects such as fictitious forces.

General relativity tells us that we accelerate with respect to the gravitational field.
In Minkowski spacetime, geodesic trajectories appear as straight lines in inertial frames. If we apply a diffeomorphism to it that takes the inertial frame to a uniformly accelerated one, the metric field transforms in a way that geodesics are now parabolas.
All effects of inertia are being experienced because the gravitational field determines particle movements (in the absence of other fields).

General covariance takes away any objectivity from spacetime on its own. What we observe from it are the effects of the gravitational field given by the metric field.
For the metric field to be able to interact, it must be a dynamical field.
We recognise that the metric is not a background anymore, general relativity is now background independent.
From this insight, it is a feasible task to find its equations of motion. As mentioned in section \ref{diffeocovariance}, diffeomorphism covariance is a strong constraint on the form of field equations.
After realising the need for it, Einstein was able to find the final equations, as sketched below.

The source equations for the Newtonian potential $V$ of gravity can be written as:
\[\Delta V = - 4 \pi G \rho\]
$\Delta$ is the Laplace operator, $G$ the gravitational constant and $\rho$ the energy density. To generalise these equations, the left side must be replaced by some geometric entity that reduces to $\Delta V$ in the static, nearly flat limit and the right hand side must be generalised by a generally covariant quantity.
The latter generalisation is a quite straightforward educated guess: The energy-momentum tensor $T^{ab}$ takes the position of $\rho$.
Replacing the left hand side is a bit harder, but still feasible if one bears in mind that the divergence of the energy-momentum tensor vanishes. One relatively simple divergence-free geometric quantity is the Einstein tensor $G_{ab}$, a linear combination of the Ricci tensor $R_{ab}$, the curvature scalar $R$ and the metric tensor $g_{ab}$:\footnote{Both Ricci tensor and curvature scalar are contractions of the Riemann curvature tensor $R^a_{bcd}$: $R_{ab} = R^c_{acb}$ and $R = R^a_a$.}
\[G_{ab} = R_{ab} - \frac{1}{2} R g_{ab}\]
Comparing the static limit to Newtonian gravity and adjusting the numerical factors leads to the Einstein equations:
\[G_{ab} = 8 \pi G T_{ab}\]
It is also possible to arrive at the field equations even faster if one guesses an action for it.
There is one simple Lagrangian density function we already encountered: The scalar curvature $R$.
Let us guess the Einstein-Hilbert action:
\[S_{EH}[g] := \int R \sqrt{-\det g} \dif^n x\]
It turns out that the Euler-Lagrange equations of this action are the Einstein equations\footnote{If we do not add an action for matter fields, we arrive at the vacuum Einstein equations $G_{ab} = 0$. Including a matter action and varying it with respect to the metric as well leads to the full Einstein equations we already encountered.} as well.

\subsection{Diffeomorphism covariance in quantum gravity}

\label{diffeocovarianceqg}

General relativity demonstrates excellently that in situations where no signs from experiment led to a satisfying theory, deductions from deep assumptions can lead very far.
In the search for quantum gravity, we find ourselves in such a situation.
We have certain clues such as the isotropy of the cosmic microwave background that might be explained by a theory of quantum gravity, but there is no clear lead to how exactly it should look like.
If we want to repeat the deduction strategy of general relativity, what are the assumptions we should now base our deduction on?
We should ask whether we can reuse Einstein's assumptions and carry them even further to a quantum theory.\\

One of his central achievements was recognising that gravity is a geometric effect: The metric must be interpreted as a dynamical field and deserves field equations.
To arrive at the Einstein equations, two essential ingredients suffice:
The Einstein-Hilbert action and diffeomorphism invariance.
%
Let us investigate these two ingredients and ask whether a theory of quantum gravity should build upon them.

\paragraph{Einstein-Hilbert action}

As we saw, the Einstein-Hilbert action is the integral over the scalar curvature $R$.
The scalar curvature is the only nonzero scalar one can arrive at through contractions and summations from the Riemann tensor to first order, so the Einstein-Hilbert action is quite a simple choice. 

Popular modifications to it include a volume term (the cosmological constant), functions of $R$ (the so-called $f(R)$-models) or quadratic contractions of the Riemann tensor (for example the Gau\ss{}-Bonnet-term).
Experiments clearly favour the Einstein-Hilbert action plus only a cosmological constant and no other terms.
However, we do not know any 
arguments to choose this action over another. We only have to demand that the Lagrangian density is diffeomorphism invariant, so the field equations are as well, but there are many invariant actions.
For example, modifications that are only visible at high energies are not excluded.
So in quantum gravity, we should not insist on the Einstein-Hilbert action but we should be open to high energy corrections.

\paragraph{Diffeomorphism covariance} 

Diffeomorphism covariance on the other side is a deep philosophical principle as we have seen.

Often, it is seen as an obstruction to quantisation, an unwieldy gauge symmetry. Attempts are being made to ``get rid of it'', to invent a picture where it does not appear at all which then is the fundamental description.
In doing so, one has to take caution with unmotivated assumptions.
If we follow such ideas, which can be approximately paraphrased as ``emergent diffeomorphism covariance'', we should ensure that it is either experimental facts or convincing arguments that force us to give up diffeomorphism covariance.
In absence of these, it is more beneficial to take diffeomorphism covariance as a key lead to the right theory - in the same manner Einstein did.

If we are forced to give up the concept of a manifold for spacetime for something else, we have to ask if this ``quantum spacetime'' admits a corresponding symmetry that augments diffeomorphism covariance.\\


Independently, we should also insist on a nonperturbative definition.
Especially in quantum gravity, we expect interesting effects to occur in black holes or the big bang, where the curvature becomes arbitrarily large. A perturbative expansion around a fixed background will not describe such situations properly.

\section{String theory}

Perturbative string theory originated from the study of strong interactions. 
When it was discovered that in closed string theory excitations appear that have the same propagator as a graviton (the first order of perturbation of classical gravity around Minkowski spacetime), the idea that string theory might be capable of describing quantum gravity was born.

Nowadays, string theory is developed far beyond perturbation theory. There are many alternative attempts to define it, such as the AdS/CFT-correspondence or string field theory, which often do not translate completely into each other.
An in-depth account of all these formulations is far beyond the scope of this essay.
We will only briefly touch perturbative string theory, explain its discrepances with background independence and then look at a popular possible resolution, the AdS/CFT-correspondence.
A recent overview of string field theory can be found in \cite{approachestoqg:taylor}.

\subsection{Perturbative String theory}

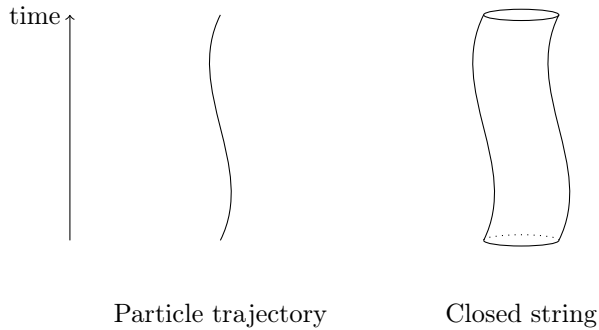
\begin{figure}[h]
	\begin{center}
		\begin{tikzpicture}[yscale=0.5, every text node part/.style={align=center},node distance=0.6cm]
			\draw[->] (-5,-3) -- +(0,6) node[left] {time};
			\foreach \x/\wann in {-3cm/1,0.5cm/2,1.5cm/2}
			{
				\begin{scope}[xshift=\x]
					\draw (0,-3) .. controls (0.5,-1) and (-0.5,1) .. (0,3);
				\end{scope}
			}
			\draw (1,3) ellipse (0.5cm and 0.15cm);
			\draw[dotted] (1.5,-3) arc (0:180:0.5cm and 0.15cm);
			\draw (0.5,-3) arc (180:360:0.5cm and 0.15cm);
			\node (p) at (-3,-5) {Particle trajectory};
			\node (s) at (1,-5) {Closed string};
		\end{tikzpicture}
	\end{center}
	\caption{Trajectories in string theory and point particle physics} \label{figstring}
\end{figure}

An easy intuition of perturbative string theory arises by a comparison to particle physics. It is illustrated in figure \ref{figstring}.
A particle trajectory in classical, relativistic mechanics is a function $x: \R \to M$, where $M$ is a spacetime obeying the Einstein equations. 
The domain $\R$ is the proper time of the particle. Spatially, a particle is zero-dimensional. We can emphasise this by writing $\R \cong \R \times \{\cdot\}$, the $\cdot$ representing the point at any instance of time.

In string theory, this is generalised to a spatially one-dimensional object, a string. Now, we already have the choice of two different topologies for one-dimensional compact manifolds: An interval or a circle line. The string theory corresponding to the first choice is called ``open string theory'', the second choice is ``closed string theory''. We will focus on closed string theory.

In analogy to particle mechanics, a closed string trajectory is a function $X: \R \times S^1 \to M$, 
where $S^1$ is the circle line, the shape of a closed string to a given instance of proper time.
Also in analogy to the point particle with the proper time as the free action, a typical action chosen for the bosonic\footnote{Nowadays, the bosonic case is superseded by the supersymmetric one. The bosonic theory contains tachyonic excitation modes which might be related to instabilities of the vacuum state. Furthermore, fermions are only predicted by supersymmetric string theory.} closed string theory is the Polyakov action, which is classically equivalent to the area of the string worldsheet.\footnote{The worldsheet is the two-dimensional analogon of the one-dimensional trajectory of a point particle.} Defined in this way, the Polyakov action depends on the geometry of the manifold that the string lies in.\\

The function $X$ is now to be quantised perturbatively.
A solution to its equation of motion is just the shape of an infinitely extended cylinder. This is in accordance with particle mechanics: An observer who cannot resolve distances smaller than the string radius will see a point particle following a straight line.
If we now study excitations of the closed string around this tubular shape, we arrive two interesting results:
One of the excitation modes has a propagator exactly like the propagator one would expect for a graviton.
Furthermore, for the quantisation to be consistent, it turns out that we have to demand the metric on $M$ to satisfy certain differential equations, some of which reduce to the Einstein equations with higher order corrections.
We therefore directly recover general relativity as an effective, classical limit and can interpret the closed, massless string excitations as perturbations around the background metric on $M$.
Together with the conjectured UV-completeness, these two results are a strong hint that closed perturbative string theory might lead the way to a candidate theory of quantum gravity.

However, the perturbative approach cannot be the final, fundamental answer for the reasons discussed in \ref{perturbative}:
It will not make meaningful predictions in cases where we want to allow the geometry to fluctuate strongly, including the interesting cases of black holes and the big bang.

\subsection{AdS/CFT}

The aim of this subsection is not to give a detailed report of the AdS/CFT-correspondence, but rather highlight its defining features in a non-technical manner in order to study its background independence.\\

Dualities play an important role in string theoretical research.\footnote{This is not the only place where they arise. For example in quantum field theory, in the study of solitons, dualities also play an important role.} A duality exists if there are two different mathematical descriptions of the same physical theory. Dualities arise together with gauge redundancies: The two descriptions are different because they add different gauge redundancies. They only differ by unphysical content.

A simple prototype for such a duality might be two gauge theories (not necessarily Yang-Mills) that have different configuration spaces, but after identifying gauge equivalent configurations, the physical state spaces are isomorphic.
This is quite a strong duality, and the study of dualities is not limited to this cases.
A weaker type of duality occurs if we demand that predictions of the values all observable quantities always agree.

Seiberg discussed dualities where two perturbative descriptions both are weak coupling limits of the same ``full'' theory, but with different couplings. For example, the first theory is a good approximation to the full theory whenever some coupling constant $g$ is small, whereas the second theory appears in the same manner whenever $\frac{1}{g}$ is small. This type of duality is called ``S-duality''.
In such a situation, the first theory can be used to study the effects of the second theory when it is strongly coupled, and vice versa. We can for example use perturbative methods in the first theory to derive nonperturbative results of the second theory.
A compact discussion of this situation is found in \cite{rozali}.\\

The Anti-de Sitter/Conformal Field Theory conjecture (due to Maldacena in 1997) is a duality between two very different theories.

The first part is supersymmetric string theory in asymptotic Anti-de Sitter spacetime, which will be abbreviated by AdS.
Anti-de Sitter spacetimes are homogeneous and isotropic solutions to the Einstein equations with negative cosmological constant. Because of their high symmetry, an AdS spacetime is determined by its cosmological constant. Furthermore, every AdS spacetime is related to every other AdS spacetime by a global conformal transformation. Therefore, it is usual to speak about ``the'' AdS spacetime.

The second part is supersymmetric $SU(N)$ - Yang-Mills theory on the boundary of AdS.\footnote{The AdS/CFT-correspondence is generalised by the ``gauge/gravity'' or ``gauge/string'' duality conjecture, which is not stated precisely. One hopes for a similar duality like AdS/CFT to be true for other asymptotic geometries than AdS (the gravity part) and other boundary gauge field theories (the gauge part). This is in line with the holographic principle suggested by 't Hooft and Susskind in the nineties.} It is maximally supersymmetric with four supersymmetries.
For the duality to become apparent, the number of colours $N$ will tend to infinity, while at the same time the 't Hooft coupling $g^2N$ stays constant, where $g$ is the coupling of the Yang-Mills theory. The value of the 't Hooft coupling decides whether the Yang-Mills theory is approximated well by perturbation theory or not.\\

\paragraph{Sketch of the duality}

An asymptotic AdS geometry can be quite arbitrary away from the boundary, but must approximate AdS geometry at the asymptotic boundary of AdS on the boundary. As explained in \cite[IV A]{rozali}, it is generally believed that there are no local, gauge invariant quantities in general relativity and therefore a partition function for the string fields encoding the metric does not have source terms, but only boundary conditions, which we will formally write as $J$.
Any observables we might probe in an experiment, such as radiation coming from a black hole or gravitons being emitted from a highly energetic process should be calculable by means of the partition function $Z_{\text{AdS}}[J]$.

Since the boundary conditions $J$ are localised on the boundary $\partial$AdS, we can now ask whether the partition function can be written entirely in terms of quantities on the boundary.
This seems indeed to be the case and is conjectured to be true exactly.
The corresponding description is the partition function of $SU(N)$ - Yang-Mills theory in the limit of large $N$ with four supersymmetries, where $J$ now plays the role of a genuine source term.
This specific formulation of the AdS/CFT-conjecture in the fashion of Witten (see \cite{witten}), could be stated in means of formulas roughly as:
\[Z_{\text{AdS}}[J] := \text{\LARGE ``}\int\limits_{\mathrlap{g|_{\partial\text{AdS}} \approx J}} Dg e^{iS[g]}\text{ \LARGE ''} \stackrel{\text{conjectured}}{=} Z_{\text{CFT}}[J] := \int DA e^{iS_{\text{YM}}[A,J]}\]
The left hand side is in quotation marks since there is no accepted rigorous definition of this partition function.

One first result is the radius of AdS. It turns out to be proportional to the fourth root of the 't Hooft - coupling and the typical radius $l_s$ of a string: $R_{\text{AdS}} \propto \sqrt[4]{g^2N} l_s$.
This tells us that we are in the situation of an S-duality: For large 't Hooft couplings, spacetime is much larger than the strings it contains, so string perturbation theory becomes a good approximation.
For small 't Hooft couplings, the Yang-Mills theory can be perturbed itself.
If the duality is true in general, 

\subsubsection{Background independence of the AdS/CFT-approach}

Instinctively, one identifies the constraint of the geometry being asymptotically AdS as a background and concludes that AdS/CFT is background dependent by construction.

However, it has been argued that the situation is different:
In classical general relativity, there is no process that starts with asymptotic AdS and ends with another asymptotic geometry as for example the flat geometry.
If one accepts the idea that changing the asymptotic geometry is an impossible, nonphysical process, one should only consider physical states that are related to the current state by a physical process. Such states are then said to belong to the same ``superselection sector''.
Choosing a particular superselection sector then is no longer a physical constraint, but it should rather be interpreted as choosing a particular parameter\footnote{This argument is similar to the one in section \ref{diffeocovariance} where we discussed reinterpreting constant background fields as constants of nature.} of the theory.

A superselection sector is the opposite of a background: It appears in the states of the theory, but there is no physical process that can change it.
A background can be changed by a physical process, but our theory does not allow for such a change, whereas there exists a fundamental theory that allows for it.

In this sense, constraining the asymptotic geometry is not a background dependence.\\

We also realise that not all diffeomorphisms now are allowed gauge transformations: A diffeomorphism must leave a neighbourhood around the boundary identical in order not to transform the state into another superselection sector.

These diffeomorphisms are commonly called ``bulk diffeomorphisms'', where ``bulk'' refers to everything related to the spacetime that does not influence the boundary.
It seems clear that the state of the gauge theory on the boundary does not change under a bulk diffeomorphism.
So if the duality between the two theories can indeed be established, the gauge theory guarantees diffeomorphism invariance.
By duality, the string theory in asymptotic AdS also respects diffeomorphism invariance.\footnote{In the sense that diffeomorphisms are gauge transformations.} 

If the AdS/CFT conjecture is true, background independence of string theory could be proven, following this argument.

\subsubsection{Open and partially solved problems}

The major problem is the lack of an actual proof of the conjecture.
Exactly the reason why S-dualities are useful often makes them hard to prove: Whenever one of the descriptions is in the well-behaved, weakly coupled regime, the other is strongly coupled and usually intractable.

There are good arguments in favour of the conjecture and there have been a lot of qualitative and quantitative examples where the two descriptions agree. 
No counterexamples have yet been discovered.
An actual proof is missing, however.

Part of the problem is also the AdS part of the duality being ill-defined. String theory can be defined perturbatively, but as argued before this cannot be the fundamental definition, which is still lacking. Attempts to define it with e.g. string field theory have not succeeded yet.

It has been argued that the AdS/CFT duality might be the best way to actually define string theory: It would be whatever is dual to the gauge theory on the boundary.
The duality does not become a tautology in this scenario. Many important features of string theory can still be calculated with the perturbative definition. Every nonperturbative definition should agree with it in the regime of small perturbations.\\

The main problems regarding background independence are these related ones:

If we want to perturb the theory, we have to choose a vacuum state which is asymptotically AdS. Regardless of background independence, that is, of our ability to chose any such vacuum, this is a strong restriction:
We are bound to a particular superselection sector. If we want to change it, we would have to find another duality. This is a conceptual problem, but it has also implications if we want to achieve an agreement between string theory and experiment:
The superselection sector of AdS/CFT is not the one we are living in since the cosmological constant was measured to be positive.
If there were a duality to string theory in asymptotic de Sitter spacetime, a vacuum solution with \emph{positive} cosmological constant, this would be a much more realistic model.
This is being studied, see for example \cite{strominger:dscft}.
Unfortunately, this has not been achieved yet for technical reasons.
Likewise, we cannot apply the duality to the big bang, as the geometry departs from asymptotic AdS around the singularity.

\section{Loop quantum gravity and state sum models}

\subsection{Loop quantum gravity}

\subsubsection{Holonomies}

The conceptual starting point of loop quantum gravity is completely different to the one of perturbative quantum field theory.
The fundamental quantities to be quantised are not the values of the metric at different points in spacetime, but so-called holonomies.

Generally, a ``holonomy'' is a transformation of a space attached to a point on a manifold to the space of another point on the same manifold. A so-called ``connection'' assigns a holonomy to every path on the manifold.\\

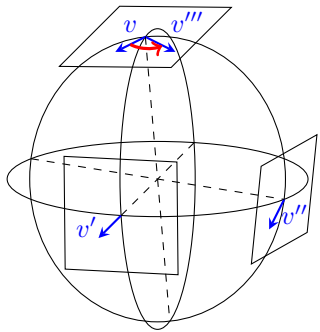
\begin{figure}[h]
	\begin{center}
			\begin{tikzpicture}[vector/.style={->,thick,>=stealth,blue}]
		\draw [name path = ell1] (0,0) ellipse (1.7cm and 1.9cm);
		\draw [name path = ell2] (0,0) ellipse (0.5cm and 2cm);
		\draw [name path = ell3] (0,0) ellipse (2cm and 0.5cm);

		\path [name intersections={of=ell1 and ell2}];
		\draw[vector] (intersection-2) -- node[above=1pt] {$v$} +(-0.4,-0.2);
		\node[rectangle, xslant=1, yslant=-0.01, minimum height=0.8cm, minimum width=1.5cm, draw] at (intersection-2) {};
		\draw[vector] (intersection-2) -- +(0.4,-0.2);
		\node[blue] at ($(intersection-2)+(0.56,0.2)$) {$v'''$};
		\draw[red, very thick, ->] ($(intersection-2)+(-0.2,-0.1)$) arc (220:310:0.3cm and 0.15cm);
		\draw[dashed] (intersection-2) -- (intersection-4);
		
		\path [name intersections={of=ell2 and ell3}];
		\draw[vector] (intersection-3) -- node[left=1pt] {$v'$} +(-0.3,-0.3);
		\node[rectangle, xslant=-0.01, yslant=-0.05, minimum size=1.5cm, draw] at (intersection-3) {};
		\draw[dashed] (intersection-1) -- (intersection-3);

		\path [name intersections={of=ell1 and ell3}];
		\draw[vector] (intersection-4) -- node[right=-2pt] {$v''$} +(-0.2,-0.4);
		\node[rectangle, xslant=0.1, yslant=0.6, minimum height=1.3cm, minimum width=0.7cm, draw] at (intersection-4) {};
		\draw[dashed] (intersection-2) -- (intersection-4);
	\end{tikzpicture}
	\end{center}
	\caption{The holonomy from the north pole to Africa, Indonesia and back to the north pole is a rotation by a right angle}\label{holonomy}
\end{figure}

Figure \ref{holonomy} illustrates the geometric intuition of a holonomy with an example:
First, we choose a vector in the tangent space of the north pole.
So in this example, the manifold is the surface of the earth, the starting point on the manifold is the north pole and the attached space is the tangent space.
We then transport the vector $v$ parallely along the quarter of a great circle down to the Democratic Republic of Congo, ending up with a vector $v'$.
Transporting vectors parallely along curves is a connection, the so-called Levi-Civita connection, which is the connection used in General Relativity.
We can go on a further path from Congo to Indonesia, resulting in $v''$ and back to the north pole, finding that the final vector $v'''$ is the original vector $v$ rotated around a right angle.
So this rotation is the holonomy of the closed loop.

The holonomy depends on the path chosen. The holonomy of a closed loop quantifies the amount of curvature in the interior of the loop. So it contains significant information about the geometry of the manifold.
Indeed, given the holonomies of all paths, it is possible to reconstruct the connection and therefore the curvature tensor completely.\\

Holonomies are chosen to be the fundamental quantities because in a specific sense, they are diffeomorphism invariant:
If we take a diffeomorphism and pull back the fields encoding the geometry (the connection and therefore the curvature) as well as the paths, then the holonomies of the paths will stay unchanged.
It has to be noted here that paths of course are not fields. We are leaving the realms of classical field theory again and had to augment the definition of the action of a diffeomorphism in such a way that it now respects the paths as dynamical variables.
This is in agreement with the resolution of the hole argument we discussed in \ref{holeargument}.
There is no intrinsic meaning to points on the manifold in themselves. Pulling back fields and leaving the points of a path at their place is inconsistent.

\subsubsection{Spin networks}

The alert reader may have noted that a connection usually does not contain all of the geometric information. It might seem wrong to just quantise just a connection.
However, in the Hamilton-Jacobi formalism of general relativity as described in \cite[Section 4]{rovelli:qg}, one finds that the gravitational field can be described by two conjugate variables. One of them is an $SU(2)$-connection,\footnote{The transformations associated to paths are actions of $SU(2)$ elements.}  the other one contains information about the spatial metric and behaves like an associated momentum to position.
When quantising just the connection, the other variable does not need a separate Hilbert space.

It is therefore enough to understand how to quantise $SU(2)$ and to understand how gauge transformations act on the Hilbert space.
$SU(2)$ appears as the space of holonomies on a graph whose vertices are points of the manifold and the edges are paths on the manifold.
The gauge transformations in question are the $SU(2)$-transformations of the $SU(2)$-connection, which we will not discuss here, and diffeomorphisms.\footnote{It turns out that the ordinary group of diffeomorphisms has to be extended for technical reasons as explained in \cite[6.4.2 and 6.7]{rovelli:qg}.}

There is a plethora of different graphs on a manifold. The key insight now is the following: Given two graphs that are isomorphic as graphs, but embedded at different places on the manifold, then we may find a diffeomorphism that maps the first graph on the second, if they are not knotted in a different way. Since diffeomorphisms are gauge transformations, we identify both these graphs as the same physical state if they have the same holonomies.
So we do not need to care about how a graph is embedded in a manifold\footnote{If one insists, this is a generalisation of the hole argument as described before: There is no physical meaning of the embedding of the graph, in itself.} but just deal with the abstract graph and the holonomies on it.

If we now consider the Hilbert space over abstract graphs\footnote{Actually, two embedded graphs can only be mapped one onto the other if both graph embeddings are in the same ``knot equivalence class''. So in fact we have to count different graphs and different knots. This does not make the Hilbert space untractably large.} with holonomies, we find that it is spanned by so-called ``spin networks''. 
There are a lot of details left out in this overview. Also since there is much progress happening in loop quantum gravity, the exact quantisation procedure may be different depending on which specific approach one follows.

A spin network is a graph together with a labelling of irreducible unitary representations of $SU(2)$ on the edges and equivariant maps (so-called ``intertwiners``) of the representation spaces on the vertices.
The irreducible unitary representations of $SU(2)$ can be counted by spins, that is, half integers.
This is the origin of the name ''spin network`` which was coined by their inventor, Roger Penrose.

It can be shown that if an edge is labelled with a spin $j$, then there is a triangle pierced perpendicularly by the edge which has area proportional to $\sqrt{j(j+1)}$ Planck areas.
Other geometric quantities such as lengths and volumes have similarly discrete formulas.
A spin network is one particular implementation of the idea of a ''quantum geometry``, or a ''quantum space``.

Figure \ref{spinnetwork} illustrates a spin network.

\tikzset{vertex/.style={circle, fill, inner sep=0mm, minimum size=1mm}, edge/.style={postaction={decorate},decoration={markings,mark=at position .55 with {\arrow{>}}}}}
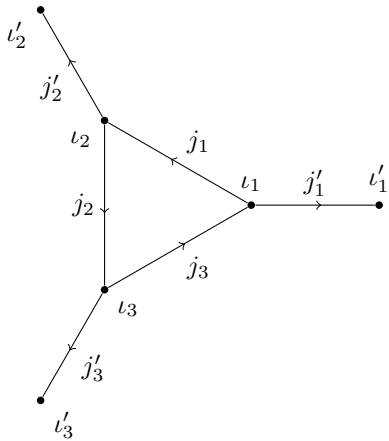
\begin{figure}[!h]
	\begin{center}
		\begin{tikzpicture}
			\foreach \angle/\anglelabel in {0/1,120/2,240/3}
			{
				\node[vertex,label=90+\angle:$\iota_\anglelabel$] (i\anglelabel) at (\angle:1.3cm) {};
				\node[vertex,label=90+\angle:$\iota_\anglelabel'$] (o\anglelabel) at (\angle:3cm) {};
				\draw[edge] (i\anglelabel) -- node[auto] {$j_\anglelabel'$} (o\anglelabel);
			}
			\foreach \ala/\alb in {1/2,2/3,3/1}
			{
				\draw[edge] (i\ala) -- node[auto,swap] {$j_\ala$} (i\alb);
			}
		\end{tikzpicture}
	\end{center}
	\caption{A spin network is a graph labelled with spins (the $j_i$) on the edges and intertwiners (the $\iota_i$) on the vertices.}\label{spinnetwork}
\end{figure}

An aesthetical advantage of this whole approach is the role played by diffeomorphisms.
Without diffeomorphisms as gauge transformations, the configuration space would consist of any embedded graphs with holonomies on them.
This space is intractably large and leads to a nonseparable Hilbert space.
After having divided by diffeomorphisms, the space is separable and has an easy to handle basis, the spin networks.
So diffeomorphism invariance is prominently made use of. It is not an obstruction, but an essential building block. Loop quantum gravity would not work without it.

\subsubsection{Problems}

From a philosophical perspective, one detail of the whole construction seems improper:
Building upon a Hamiltonian approach. This artificially separates a time direction from a space direction. A few problems arise here:

It is not obvious how find a discrete Hamiltonian, given the classical one. Much work has been done in that direction, but the results seem quite technical and complicated.

In an approach that splits space and time at the beginning, Lorentz invariance needs to be shown subsequently. This has been done already, but philosophically, an approach without such a splitting would be preferable.\\

The technical hardships of loop quantum gravity were only hinted at.
One might fall for the simplified picture that one ''just`` has to take holonomies on graphs, divide them by diffeomorphisms, quantise them and end at an easy Hilbert space spanned by spin networks.
Quite in the contrary, there are a lot of mathematical details to be taken into account before one arrives there. 
And once this is done, the theory is only \emph{defined} and not yet understood.

In this essay, there is no attempt to explain how one studies this theory and relates it to measurable quantities, as this is beyond the scope of this essay. There has not been any attempt to do this for string theory, either.
For a seminal paper on the classical limit of loop quantum gravity which still leaves technical questions open, see \cite{thiemann-giesel:classiclimit}.
It should only be noted at this point, that the reasons why the classical limit of loop quantum gravity still has not been proven completely to be general relativity are not purely technical.

One reason for the classical limit being hard to find is exactly the usage of background independent variables.
In a regular experiment, we will set up a coordinate system and try to measure quantities in that coordinate system. When measuring the effects of a quantum field theory, we might assume a vacuum configuration (a background) and try to measure small perturbations (particles) of it.
This is quite close to any theory that predicts for example a graviton directly.

By starting with holonomies, loop quantum gravity predicts the behaviour of different observables than for example perturbative quantum field theory. There is no obvious way to read off a graviton from a spin network state.
By choosing background independent observables, the classical limit is hard to take because we typically perform experiments fixing some coordinate system, as has been argued in section \ref{holeargument}.\footnote{There are of course exceptions from this observation. The E821 experiment at Brookhaven National Laboratory determined the anomalous magnetic moment by indirectly measuring the holonomy of the magnetic field of a ring accelerator. All interferometres such as gravitational wave detectors indirectly measure the differences of proper times of photons travelling on the arms. These are all examples which come more or less close to the idea of measuring a holonomy directly.}
Therefore theories not respecting diffeomorphism invariance are often closer to our intuition.

\subsection{State sum models and spin foams}

State sum models have been used in statistical mechanics and even in algebraic topology.
A particular flavour, so-called spin foam models have been devised to overcome some of the complicacies of loop quantum gravity.
In the same way as the Feynman path integral provides a Lorentz covariant approach to quantum field theory without having to specify a Hamiltonian, state sum models try to define dynamics for spin networks without the space-time-splitting and the Hamilton-Jacobi approach.

In perturbative quantum field theory, a basis of the Hilbert space is given by (point-like) particle states.
Feynman diagrams connect the incoming point particles with the outgoing ones via one-dimensional edges and zero-dimensional vertices whose valency depends on the action.
Now spin network states already consist of vertices and edges, so consequently, the ''spin foams``, which connect spin network states, consist of vertices, edges and two-dimensional faces.
Like Feynman diagrams give rise to transition amplitudes between particle states, one can calculate transition amplitudes of spin networks from spin foams.
Like spin networks can be interpreted as ''quantum spaces``, spin foams can be viewed as ''quantum spacetimes``.

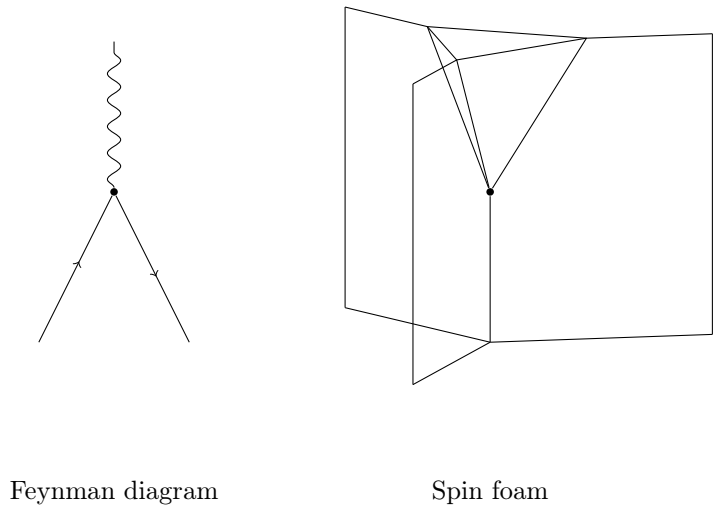
\begin{figure}[!h]
	\begin{center}
		\begin{tikzpicture}[z={(0,2)},y={(0,0.2)}]
			\begin{scope}[every node/.style={coordinate}]
				\node (ma) at (0,0) {};
				\node[vertex] (mm) at (cylindrical cs:angle=0,radius=0,z=1) {};
				\draw (ma) -- (mm);
				\foreach \angle/\vertexlabel in {0/1,120/2,240/3}
				{
					\node (ia\vertexlabel) at (cylindrical cs:angle=\angle,radius=3,z=0) {};
					\draw (ma) -- (ia\vertexlabel);
					\node (ib\vertexlabel) at (cylindrical cs:angle=\angle,radius=1.3,z=2) {};
					\node (ic\vertexlabel) at (cylindrical cs:angle=\angle,radius=3,z=2) {};
					\draw (mm) -- (ib\vertexlabel) -- (ic\vertexlabel);
					\draw (ia\vertexlabel) -- (ic\vertexlabel);
				}
			\end{scope}
			\foreach \vla/\vlb in {1/2,2/3,3/1}
			{
				\draw (ib\vla) -- (ib\vlb);
			}
			\node at (0,0,-1) {Spin foam};
			\begin{scope}[shift={(-5,0,0)}]
				
				\node[vertex] (g) at (0,0,1) {};
				\coordinate (photon) at (0,0,2);
				\coordinate (fout) at (1,0,0);
				\coordinate (fin) at (-1,0,0);
				\draw[edge] (fin) -- (g);
				\draw[edge] (g) -- (fout);
				\draw[decorate, decoration={snake}] (g) -- (photon);
				\node at (0,0,-1) {Feynman diagram};
				
			\end{scope}
		\end{tikzpicture}
		\caption{A spin foam relates to a spin network as Feynman diagrams relate to particle states.}
		\label{spinfoam}
	\end{center}
\end{figure}

An example is shown in figure \ref{spinfoam}.
In the Feynman diagram, a fermion and an antifermion form the initial state. They annihilate to a photon, which forms the final state.
In the spin foam, a star-shaped spin network enters. The center of the star splits up to a triangle. The final state thus is the spin network from the previous figure \ref{spinnetwork}.

\makeatletter
\define@key{cylindricalkeys}{angle}{\def\myangle{#1}}
\define@key{cylindricalkeys}{radius}{\def\myradius{#1}}
\define@key{cylindricalkeys}{z}{\def\myz{#1}}
\makeatother
\tikzdeclarecoordinatesystem{cylindrical}%
{%
	\setkeys{cylindricalkeys}{#1}%
	\pgfpointadd{\pgfpointxyz{0}{0}{\myz}}{\pgfpointpolarxy{10+\myangle}{\myradius}}
}

One important difference to the Feynman path integral has to be noted:
While Feynman diagrams can be 
derived from an integral over all field configurations, it is still being debated whether such a derivation can be given for spin foams.
Instead, spacetime is being triangulated manually.

It turns out that some state sum models (mainly those modelled after topological field theories) predict the same transition amplitudes independently of the triangulation chosen.
However, in all physically relevant models up to now, different triangulations can lead to different amplitudes.

This is often raised as criticism.
It is believed by some that triangulation independence is the discrete version of diffeomorphism invariance\footnote{This is in the sense of section \ref{diffeocovarianceqg}, where we remarked that if our picture of spacetime as a manifold had to be replaced by a ''quantum spacetime`` like a spin foam, we might have to replace diffeomorphism invariance by an appropriate concept.} and therefore must not be broken by a state sum model of physical relevance.
The opponents argue that diffeomorphism invariance is already guaranteed by the initial choice of observables and that triangulation independence has only to be expected from topological field theories.

It is hard to answer this question decisively. A triangulation is certainly not a background field, but we have left the realm of classical field theory long ago, already with the introduction of holonomies. It is unclear how the notion of background independence translates exactly to this situation.
According to our characterisation of the terms ''background`` and ''superselection sector``, we now have to ask whether there is a physical process that allows for a change of the triangulation, and if yes, we have to find its dynamics. If we achieve to do this, a particular triangulation at least is not a background of the theory.\\

Apart from philosophical considerations, there is some promising progress in research on state sum models. While the first four-dimensional candidate for a good model, the Barrett-Crane model (\cite{barrett-crane}), was shown not to have the expected long-range limit, a more recent model, the EPRL\footnote{Engle, Pereira, Rovelli, Livine, see \cite{eprl}.} model was shown to have correct graviton propagator.
A full proof of agreement with general relativity to all orders is yet to be worked out.

\section{Conclusion}

We have established a working vocabulary of background independence and related concepts in classical field theory.
It turned out that in our definitions, background independence is a feature that appears in every diffeomorphism covariant theory and that excludes a perturbative fundamental definition of the field theory.
We have also learned that whenever it comes to quantum theories, we have to redefine our classical vocabulary. Sometimes, there is no obvious redefinition.\\

In string theory, we saw that old-fashioned string perturbation theory certainly is not the fundamental theory of quantum gravity.
Its perturbative nature breaks background independence and at the same time makes perturbative string theory unfit for situations with high curvature such as the big bang or black holes.
On the other hand, it contains an excitation mode which can be identified with the graviton.

The AdS/CFT conjecture provides a background independent formalism of string theory. It remains unclear to what extend it can be proven or if it should be adopted as the definition of string theory instead.
The AdS/CFT approach forces us to commit to a particular superselection sector. According to cosmological observations, this is the wrong one. Whether there exists a duality for the more relevant de Sitter spacetime remains open.

We have not discussed other contemporary approaches towards a fundamental definition of string theory, such as matrix theory or string field theory. Background independence in string theory is a vast subject that has not been scrutinised in detail yet.\\

The choice of background independent quantities, the holonomies, guarantees that loop quantum gravity is itself background independent. It could be shown that the theory is invariant under the only possible background dependency, the space time splitting.

State sum models avoid the space time splitting at all and give a compelling picture of quantum spacetime.
Apart from the unresolved question whether triangulation dependence breaks diffeomorphism invariance, they are background independent as well.
Furthermore, the recent EPRL model has been shown to exhibit the graviton propagator as well.\\

Many more approaches could be examined here. Some are not oriented to the philosophy of general relativity at all, like asymptotic safety. Others, such as dynamical triangulations, are more in the spirit of loop quantum gravity.
Background independence has been discussed more in some of the approaches and less in others.
Hopefully, the reader has been convinced that it is an issue of fundamental importance.

\newpage
\nocite{approachestoqg:horowitzpolchinski}
\nocite{belot:Background-Independence}
\bibliographystyle{plain}
\bibliography{ART.bib,LQG.bib,BI.bib,Strings.bib}

\begin{thebibliography}{10}

\bibitem{barrett-crane}
John Barrett and Louis Crane.
\newblock Relativistic spin networks and quantum gravity.
\newblock J.Math.Phys. 39 (1998) 3296-3302, 1997.

\bibitem{belot:Background-Independence}
Gordon Belot.
\newblock Background {Independence}.
\newblock arXiv:1106.0920v1.

\bibitem{eprl}
Jonathan Engle, Etera Livine, Roberto Pereira, and Carlo Rovelli.
\newblock Lqg vertex with finite immirzi parameter.
\newblock {\em Nuclear Physics B}, 799:136–149, 2008.

\bibitem{approachestoqg:horowitzpolchinski}
Gary Horowitz and Joseph Polchinski.
\newblock Gauge/gravity duality.
\newblock In {\em Approaches to quantum gravity}. Cambridge University Press,
  2009.

\bibitem{rovelli:qg}
Carlo Rovelli.
\newblock {\em Quantum Gravity}.
\newblock Cambridge University Press, 2007.

\bibitem{rozali}
Moshe Rozali.
\newblock Comments on background independence and gauge redundancies.
\newblock {\em Advanced Science Letters}, 2(2):244--250, 2009.

\bibitem{strominger:dscft}
Andrew Strominger.
\newblock The ds/cft correspondence.
\newblock arXiv:hep-th/0106113v2.

\bibitem{approachestoqg:taylor}
Washington Taylor.
\newblock String field theory.
\newblock In {\em Approaches to quantum gravity}. Cambridge University Press,
  2009.

\bibitem{thiemann-giesel:classiclimit}
Thomas Thiemann and Kristina Giesel.
\newblock {Algebraic Quantum Gravity (AQG). II. Semiclassical Analysis}.
\newblock {\em Classical and Quantum Gravity}, 24:2499--2564, 2007.

\bibitem{witten}
Edward Witten.
\newblock Anti de sitter space and holography.
\newblock {\em Advances in Theoretical and Mathematical Physics},
  2(2):253--290, 1998.

\end{thebibliography}

\end{document}